\documentclass[journal=jpccck,manuscript=article, layout=traditional]{achemso}

\usepackage{chemformula} 
\usepackage[T1]{fontenc} 

\author{Takahiro Matsuoka}
\email{tmatsuok@utk.edu / TKMatsuoka08@gmail.com}
\affiliation[Gifu U]
{Department of Electrical, Electronic and Computer Engineering, Gifu University, Gifu 501-1193, Japan}
\alsoaffiliation[UTK]
{Joint Institute for Advanced Materials, The University of Tennessee, TN 37996, USA}
\author{Shu Muraoka}
\affiliation{Materials Science and Technology Division, Graduate School of Engineering, Gifu University, Gifu 501-1193, Japan}
\author{Takahiro Ishikawa}
\email{ISHIKAWA.Takahiro@nims.go.jp}
\affiliation{The Elements Strategy Initiative Center for Magnetic Materials, National Institute for Material Science, Ibaraki 305-0047, Japan}
\alsoaffiliation[KYOKUGEN]
{Center for Science and Technology under Extreme Conditions, Graduate School of Engineering Science, Osaka University, Osaka 560-8531, Japan}
\author{Ken Niwa}
\affiliation[Nagoya U]{Department of Materials Physics, Nagoya University, Nagoya 464-8603, Japan}
\author{Kenji Ohta}
\affiliation[TITEC]{Department of Earth and Planetary Sciences, Tokyo Institute of Technology, Tokyo 152-8550, Japan}
\author{Naohisa Hirao}
\affiliation[JASRI]
{Japan Synchrotron Radiation Research Institute, Hyogo 679-5198, Japan}
\author{Saori Kawaguchi}
\affiliation[JASRI]
{Japan Synchrotron Radiation Research Institute, Hyogo 679-5198, Japan}
\author{Yasuo Ohishi}
\affiliation[JASRI]
{Japan Synchrotron Radiation Research Institute, Hyogo 679-5198, Japan}
\author{Katsuya Shimizu}
\affiliation[KYOKUGEN]
{Center for Science and Technology under Extreme Conditions, Graduate School of Engineering Science, Osaka University, Osaka 560-8531, Japan}
\author{Shigeo Sasaki}
\affiliation[Gifu U]
{Department of Electrical, Electronic and Computer Engineering, Gifu University, Gifu 501-1193, Japan}

\title[NaCl(H$_2$)]
  {Hydrogen-Storing Salt NaCl(H$_2$) Synthesized at High Pressure and High Temperature}

\keywords{hydride, NaCl, DAC, Raman, XRD, genetic algorithm technique, first-principles calculations}

\begin{document}


\begin{abstract}
X-ray diffraction and Raman scattering measurements, and first-principles calculations are performed to search for the formation of NaCl-hydrogen compound. When NaCl and H$_2$ mixture is laser-heated to above 1500 K at pressures exceeding 40 GPa, we observed the formation of NaClH$_{\textit{x}}$ with \textit{P}6$_3$/\textit{mmc} structure which accommodates H$_2$ molecules in the interstitial sites of NaCl lattice forming ABAC stacking. Upon the decrease of pressure at 300 K, NaClH$_{\textit{x}}$ remains stable down to 17 GPa. Our calculations suggest the observed NaClH$_{\textit{x}}$ is NaCl(H$_2$). Besides, a hydrogen-richer phase NaCl(H$_2$)$_4$ is predicted to become stable at pressures above 40 GPa. 
\end{abstract}

\section{INTRODUCTION}
Compressed hydrides have been predicted to be potential candidates for high-temperature (\textit{T}$_c$) superconductivity.\cite{Feng2006,Peng2017,Li2014}
Recent discoveries of metallic SH$_3$ and LaH$_{10}$ exhibiting \textit{T}$_c$ above 200 K\cite{Drozdov2015,Somayazulu2018a,Drozdov2018a} are attracting significant interest in novel hydrides. 
The synthesis of new hydrogen compounds, including unconventional stoichiometries, is of great interest.

At ambient pressure, sodium chloride (NaCl) is the only known binary compound formed by Na and Cl. 
The large electronegativity difference between Na and Cl stabilizes the highly ionic compound (Na$^+$ and Cl$^-$) which crystallizes in a rock-salt (B1) structure. 
Although it is not widely known, NaCl is one of the few materials\cite{Ohta2015,Sakamaki2009,Donnerer2013} that have been tested and revealed to be inert to hydrogen (H$_2$).
Even in a high-pressure H$_2$ atmosphere reaching 100 GPa, there is no chemistry.\cite{Ohta2015,Sakamaki2009}
NaCl does not react with H$_2$ either, when heated to 500 K at 20 GPa.\cite{Sakamaki2009}
This is why NaCl has been used to create an H$_2$-sealing capsule in high-pressure experiments.\cite{Ohta2015,Sakamaki2009,Matsuoka2011c}
However, we found the interstitial site in a B1 structure using the Pauling ionic radius\cite{Pauling1960} is large enough to accommodate an H$_2$ molecule whose molecular size is approximately 1.5 \AA.
Further, the recent experiments have revealed that alkali-metal polyhydrides (AH$_\textit{x}$, A = Li and Na, \textit{x} > 1), containing H$_2$ molecules in their unit cells, can be synthesized by heating ionic compounds such as LiH or NaH in high-pressure H$_2$.\cite{Kuno2015,Matsuoka2017,Struzhkin2016}
Research findings from the studies of AH$_\textit{x}$ suggest that the formation of NaCl(H$_2$)$_{\textit{x}}$ may be anticipated by heating NaCl and H$_2$ together at high pressures.

In this study, we performed the laser heating on the mixture of NaCl and H$_2$ (NaCl/H$_2$ mixture) to above 1500 K at pressures.
By using Raman scattering and X-ray diffraction (XRD) measurements, we observed the formation of NaClH$_\textit{x}$ with \textit{P}6$_3$/\textit{mmc} structure having H$_2$ molecules in the interstitial sites of NaCl that formed ABAC stacking, when the NaCl/H$_2$ mixture was heated at pressures exceeding 40 GPa.
The synthesized NaClH$_\textit{x}$ is suggested to be NaCl(H$_2$) by our structure search based on a generic algorithm technique and first-principles calculations.
It is implicated that NaCl(H$_2$) is formed by the H$_2$ molecules insertion into the crystal lattice of NaCl which has CsCl-type structure. 
Upon the decrease of pressure at 300 K, NaCl(H$_{2}$) remains stable down to 17 GPa.
Besides, our calculations predicted a hydrogen richer NaCl(H$_2$)$_4$ with a monoclinic (\textit{Pm}) structure was stable at pressures above 40 GPa.

\section{METHODS}
\subsection{Experimental Methods}
To detect the formation of NaCl-hydrogen compounds and reveal their crystal structure, we performed XRD and Raman scattering measurements on the NaCl/H$_2$ mixture heated to above 1500 K at pressures between 4 and 46 GPa.
The high-pressure experiments were carried out using a diamond anvil cell (DAC) equipped with type Ia diamond anvils with culets between 0.2 and 0.3 mm.
Figure 1 shows a schematic drawing of the sample chamber in a DAC.
The 0.25 mm-thick gaskets made of Re (Aldrich, 99.98$\%$) were pre-indented to 50-60 $\mu$m-thickness, and a 150 $\mu$m-diameter hole (sample chamber) was drilled through the center of the gasket.
We loaded two NaCl (Wako, 99.95$\%$+) thin plates, each of which has a thickness  of about 10 $\mu$m, in the sample chamber and filled it with fluid H$_2$.
We used IR-lasers (SPI laser, $\lambda$ = 1070 nm and $\lambda$ = 1090 nm) to heat the NaCl/H$_2$ mixture compressed in the DAC.
Because NaCl does not absorb IR radiation, tiny chips of platinum (Pt, Nilaco, 99.95$\%$) in an experiment (Exp. 1) and gold (Au, Nilaco, 99.95$\%$) in other two experiments (Exp. 2 and 3) were supplied as laser absorbers, and the IR laser was focused on Pt or Au.
Loading the H$_2$ fluid was performed by using a cryogenic gas-loading system.\cite{Chi2011} 
We first compressed the NaCl/H$_2$ mixture at room temperature to the desired pressure and then heated the mixture to above 1500 K using an IR laser. 
We note that we obtained the same results using different laser absorbers.

During compression at room temperature, we confirmed that NaCl and H$_2$ did not react by observing the structural transformation via Raman scattering measurements. 
The heating temperature was estimated by collecting thermal radiation from the sample and analyzing it within a wavelength range of 600-800 nm to convert the radiation spectrum to a temperature in accordance with the Planck's blackbody radiation law.\cite{Ohishi2008}
Laser heating was performed at BL10XU/SPring-8 (SPI laser, $\lambda$ = 1070 nm) and Nagoya University (SPI laser, $\lambda$ = 1090 nm). 
During the laser heating at BL10XU/SPring-8, the reaction between NaCl and H$_2$ was detected by monitoring the crystal structure change of NaCl by \textit{in situ} synchrotron XRD measurements using an X-ray flat panel detector (Perkin Elmer XRD0822 CP23, 1024 $\times$ 1024 pixels, pixel size: 0.20 mm). 
After heating for several minutes or after we confirmed the change of crystal structure of NaCl by XRD, the sample was quenched to room temperature. 
After quenching, we performed Raman scattering and XRD measurements at room temperature. 
The XRD patterns at 300 K were collected using an imaging plate (RIGAKU R-AXIS IV++, 300 $\times$ 300 mm$^2$, pixel size: 0.10 mm) in a forward scattering geometry. 
The Raman scattering measurements were performed in a backscattering geometry using a triple polychromator (JASCO NR1800) equipped with a liquid-nitrogen-cooled charge-coupled device. 
Radiation of 532 nm from a solid-state laser was used for excitation. 
The focused spot size of the radiation on the samples was approximately 10 $\mu$m in diameter. 
To estimate pressure, we used the shift of the first-order Raman band spectra of the diamond anvil facing the sample with a proposed calibration.\cite{Akahama2006}

\begin{figure}
 \includegraphics[scale=1]{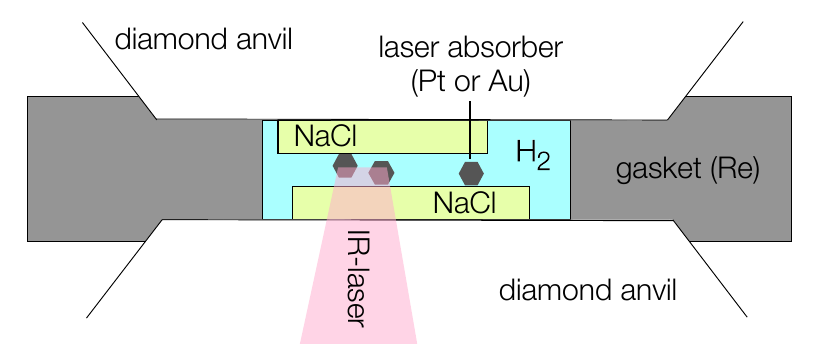}%
  \caption{Schematic drawing of the sample chamber in a DAC for the laser heating of NaCl/H$_2$ mixture.}
  \label{fgr:naclh2_1}
\end{figure}

\subsection{Computational Methods}
To investigate the possible formation of NaCl-hydrogen compounds, we performed a crystal structure search based on a genetic algorithm technique and first-principles calculations.
We developed a crystal structure prediction code based on a genetic algorithm and combined it with the Quantum ESPRESSO.\cite{Giannozzi2017,Giannozzi2009} 
We previously used this structure prediction code in the search for thermodynamically stable phases in a sulfur-hydrogen system\cite{Ishikawa2016a} and argon-hydrogen system.\cite{Ishikawa2017a} 
In our structure search, eight structures were created by `mating', six by `distortion', and six by `permutation' in each generation. 
We searched for stable structures at pressures of 30, 50, and 100 GPa using supercells including two formula units for NaClH, NaClH$_2$, NaClH$_3$, NaClH$_4$ and NaClH$_8$. 
We used the Perdew-Burke-Ernzerhof exchange-correlation functional\cite{Perdew1996} and the Vanderbilt ultrasoft pseudopotential.\cite{Vanderbilt1990}
The \textit{k}-space integration over the Brillouin zone was performed on an 8 $\times$ 8 $\times$ 8 grid, and the energy cut-off of the plane-wave basis was set to 80 Ry. 
For the phonon calculation of \textit{P}6$_3$/\textit{mmc} NaCl(H$_2$), we used an 8 $\times$ 8 $\times$ 4 \textit{k}-point grid and a 4 $\times$ 4 $\times$ 4 \textit{q}-point grid.

\section{RESULTS}
\subsection{Synthesized NaClH$_\textit{x}$}
When the NaCl/H$_2$ mixture was heated below 30 GPa, we detected no structural transformation of NaCl.
On the other hand, at 46 GPa (Exp. 1) and 42 GPa (Exp. 2 and 3) the sample exhibited visible change after laser heating (Figure 2a,b). 
The laser-focused area was distinguishable from other areas, suggesting that new material was synthesized (Figure 2b). 
The visible difference became apparent at 33 GPa (Figure 2c,d) upon a pressure decrease and revealed the presence of a compound that had a refractive index different from that of either NaCl or H$_2$. 
Hereafter, we will refer the newly formed compound as NaClH$_\textit{x}$.

\begin{figure}
 \includegraphics[scale=1]{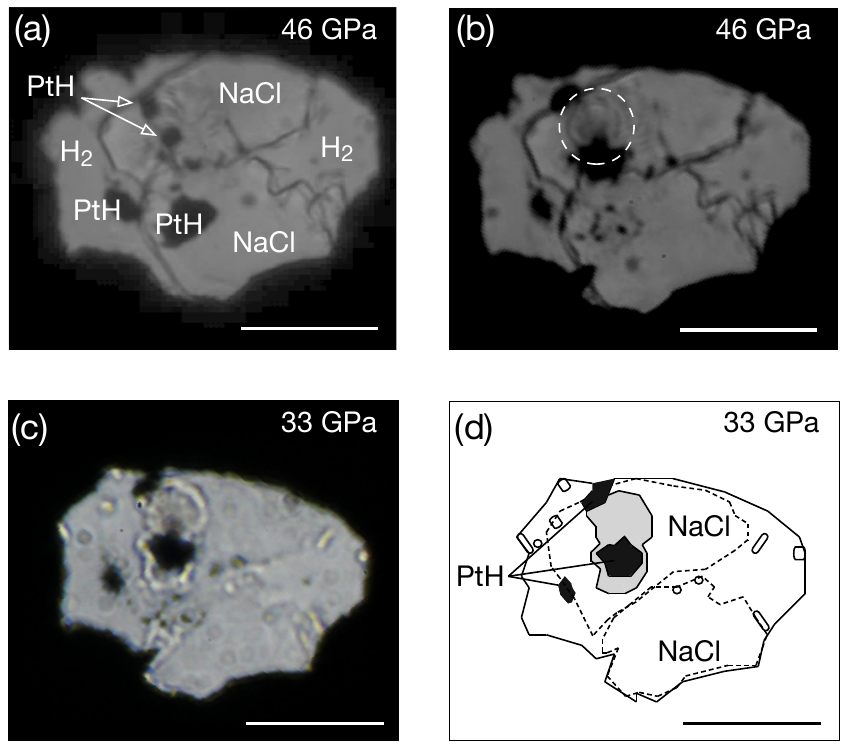}%
  \caption{Visual observation of NaClH$_{\textit{x}}$ synthesis in Experiment 1. (a) Microphotograph of the sample before laser heating at 46 GPa looking through a diamond anvil. Scale bar = 50 $\mu$m. Two pieces of PtH indicated by arrows were used for laser heating. (b) Sample after heating at 46 GPa. A dashed circle indicates the IR laser-focused area. The PtH laser absorbers moved and changed their shapes during laser heating. (c) Sample at 33 GPa upon pressure release. The bright line around the compound is Becke line. (d) Schematic drawing of the components shown in (c).}
  \label{fgr:naclh2_4}
\end{figure}

Figure 3a and b present the synchrotron XRD profiles obtained in Exps. 1 and 2. 
We note that NaCl is in a cesium-chloride-type (B2, \textit{Pm}\(\bar{3}\)\textit{m}) structure at room temperature and pressures above 29 GPa.\cite{Bassett1968,Li1987,Nishiyama2003} 
The XRD peaks of the unheated area were indexed with NaCl (B2) and PtH (\textit{P}6$_3$/\textit{mmc}). 
We did not observe the formation of gold hydride (AuH$_\textit{x}$) in Exp. 2, which is in agreement with previous reports.\cite{Donnerer2013, Ohta2015, Matsuoka2018a}
In Exp. 1 at 46 GPa after the heating, six new XRD peaks appeared (Figure  3a) in the laser-focused area in addition to the unreacted NaCl and PtH peaks. 
At 42 GPa in Exp. 2, we confirmed three additional peaks at 2\(\theta\) = 7.9$^{\circ}$, 13.8$^{\circ}$, and 17.2$^{\circ}$ (inset to Figure 3b, upper panel). 
Using these nine XRD peaks, NaClH$_\textit{x}$ was indexed with a hexagonal \textit{P}6$_3$/\textit{mmc} structure where the Cl and Na layers stack in the ABAC manner along the \textit{c} axis.
The lattice parameters were \textit{a} = 3.410(7) \AA{} and \textit{c} = 6.277(8) \AA{} at 46 GPa. 
At 20 GPa in Exp. 1 (Figure  3b, bottom panel), the presence of 010, 110 and 013 diffraction lines from the \textit{P}6$_3$/\textit{mmc} structure becomes evident due to the small overlap of the lines from  NaClH$_\textit{x}$, NaCl, and the laser absorber becomes minimal.

In Figure 3c, the volume per NaCl unit (\textit{V}$_{f.u.}$ (\AA$^3$/NaCl)), which is calculated by dividing the unit cell volume by the number of NaCl units in a unit cell, is plotted for NaCl+H$_2$ and NaClH$_\textit{x}$ as a function of pressure.
At 42 GPa, NaClH$_\textit{x}$ is 1.8\% smaller than NaCl+H$_2$ which is calculated using the XRD data of NaCl and H$_2$ observed simultaneously with NaClH$_\textit{x}$.
These observations suggest that NaClH$_\textit{x}$ is denser than NaCl+H$_2$ and energetically favorable at elevated pressures.
Before experiments, there was a concern that the laser heating of NaCl/H$_2$ might result in the formation of Na$_3$Cl, NaCl$_3$, and NaH$_\textit{x}$ (\textit{x} = 3, 7).\cite{Zhang,Struzhkin2016}
However, none of Na$_3$Cl, NaCl$_3$, and NaH$_\textit{x}$ have a \textit{P}6$_3$/\textit{mmc} lattice. 
Further, their Raman spectra do not match the data of NaClH$_\textit{x}$ (See the Supporting Information for the detailed comparisons of the Raman scattering data.).

\begin{figure}
 \includegraphics[scale=1]{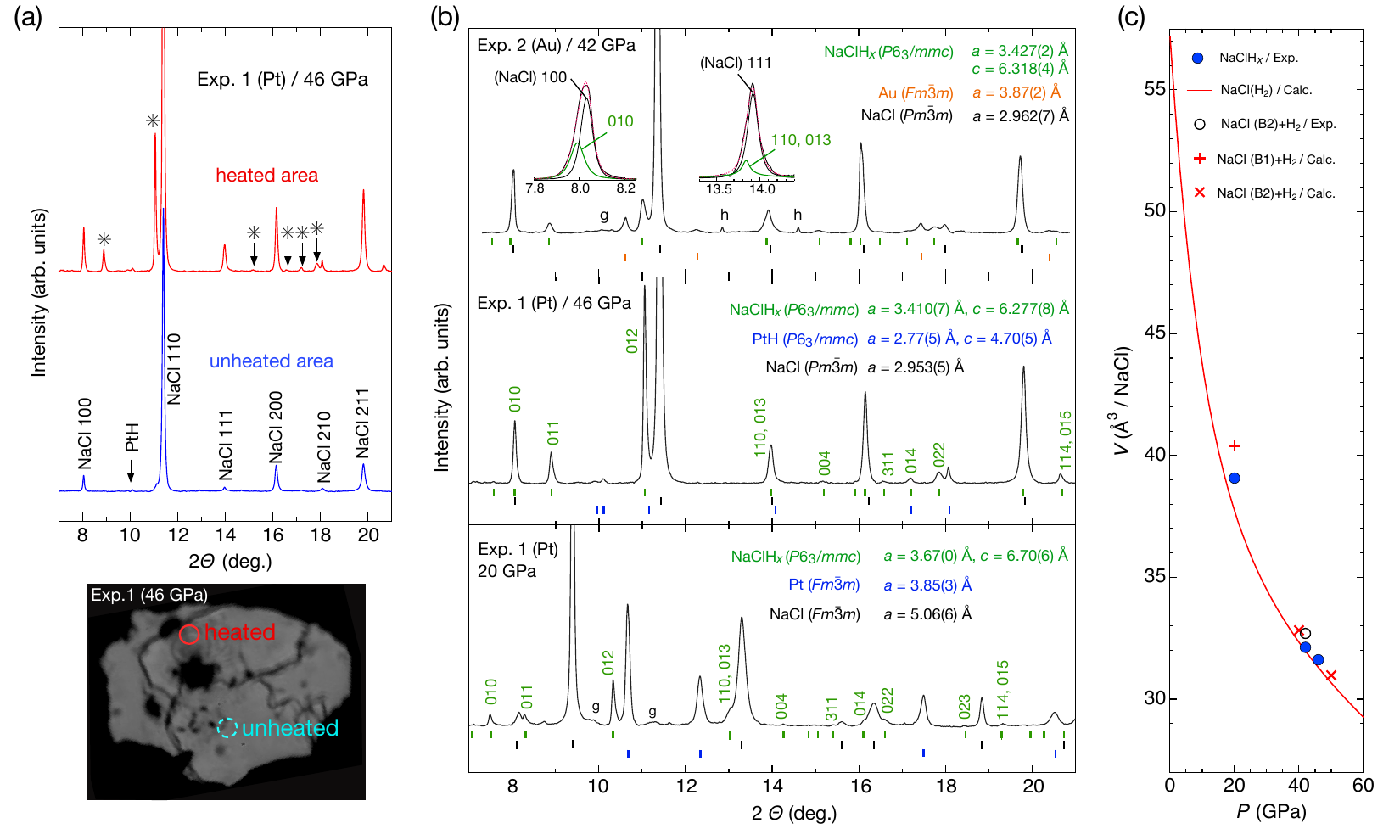}%
  \caption{XRD profiles of NaClH$_{\textit{x}}$. (a) Integrated powder XRD profiles observed at the heated and unheated areas in the sample in Exp. 1. The wavelength of the X-ray was 0.4135 \AA{} in all measurements. The asterisk (\(\ast\)) indicates peaks that appeared after the laser heating. The solid and dashed circles in the microphotograph indicate the heated and unheated areas where the XRD measurements were performed. (b) Integrated XRD profiles at pressures of 42 GPa (Exp. 2, upper panel), 46 GPa (Exp. 1, middle panel) and 20 GPa (Exp. 1, bottom panel). PtH decomposed to Pt and H$_{2}$ below 20 GPa. The notation `h' indicates the XRD peaks from solid H$_{2}$, and the `g' is the hydrogenated gasket material ReH$_\textit{x}$. (c) \textit{V}$_{f.u.}$ for the observed NaClH$_{\textit{x}}$, the predicted NaCl(H$_2$) with ABAC stacking, and NaCl+H$_2$. }
  \label{fgr:naclh2_5}
\end{figure}

Figure 4a shows the representative Raman scattering spectra measured in the heated area where examined by XRD measurements together with the data of unheated areas after the laser heating at 46 GPa. 
As the heated area was the mixture of NaClH$_{\textit{x}}$, NaCl that had no Raman-active modes, and excess solid H$_{2}$, the Raman scattering peaks from NaClH$_{\textit{x}}$ and H$_{2}$ were observed. 
Across the heated area, in addition to the rotational mode (roton) and the intramolecular stretching mode (vibron) of H$_{2}$ in excess solid H$_{2}$, three new peaks appeared at the wavenumbers 272.8, 966.4, and 4323.5 cm$^{-1}$ (hereafter referred to as \(\nu\)1, \(\nu\)2, and \(\nu\)3, respectively). 
Notably, no other new peaks appeared in the wavenumbers between 100 and 4400 cm$^{-1}$. 
The high wavenumber of \(\nu\)3 suggests that \(\nu\)3 is the vibron of H$_{2}$. 
Furthermore, the shift of \(\nu\)3 from the vibron of solid H$_{2}$ indicates that \(\nu\)3 originates from the H$_{2}$ molecules placed in a potential field different from that of solid H$_{2}$. 
Thus, NaClH$_{\textit{x}}$ contains H$_{2}$ molecules in its crystal lattice.
Figure 4b presents the Raman scattering spectra in Exp. 1 upon pressure release at 300 K. 
The Raman shift of \(\nu\)1-3 obtained in Exp. 1-3 are plotted together as a function of pressure in Figure 4c. 
Peaks \(\nu\)1-3 have positive pressure dependence.
The pressure dependences of \(\nu\)1, \(\nu\)2, and \(\nu\)3 obtained in the three experiments are in excellent agreement, indicating that the presence of Pt (PtH) or Au does not affect the results. 
When pressure is released to 17 GPa, \(\nu\)1-3 disappear, and only the roton and vibron of solid H$_2$ persist (Figure  4c). 
Because NaCl has no Raman-active phonon mode in the B1 structure, we conclude that NaClH$_{\textit{x}}$ decomposes into NaCl and H$_2$ below 17 GPa.

\begin{figure}
 \includegraphics[scale=1]{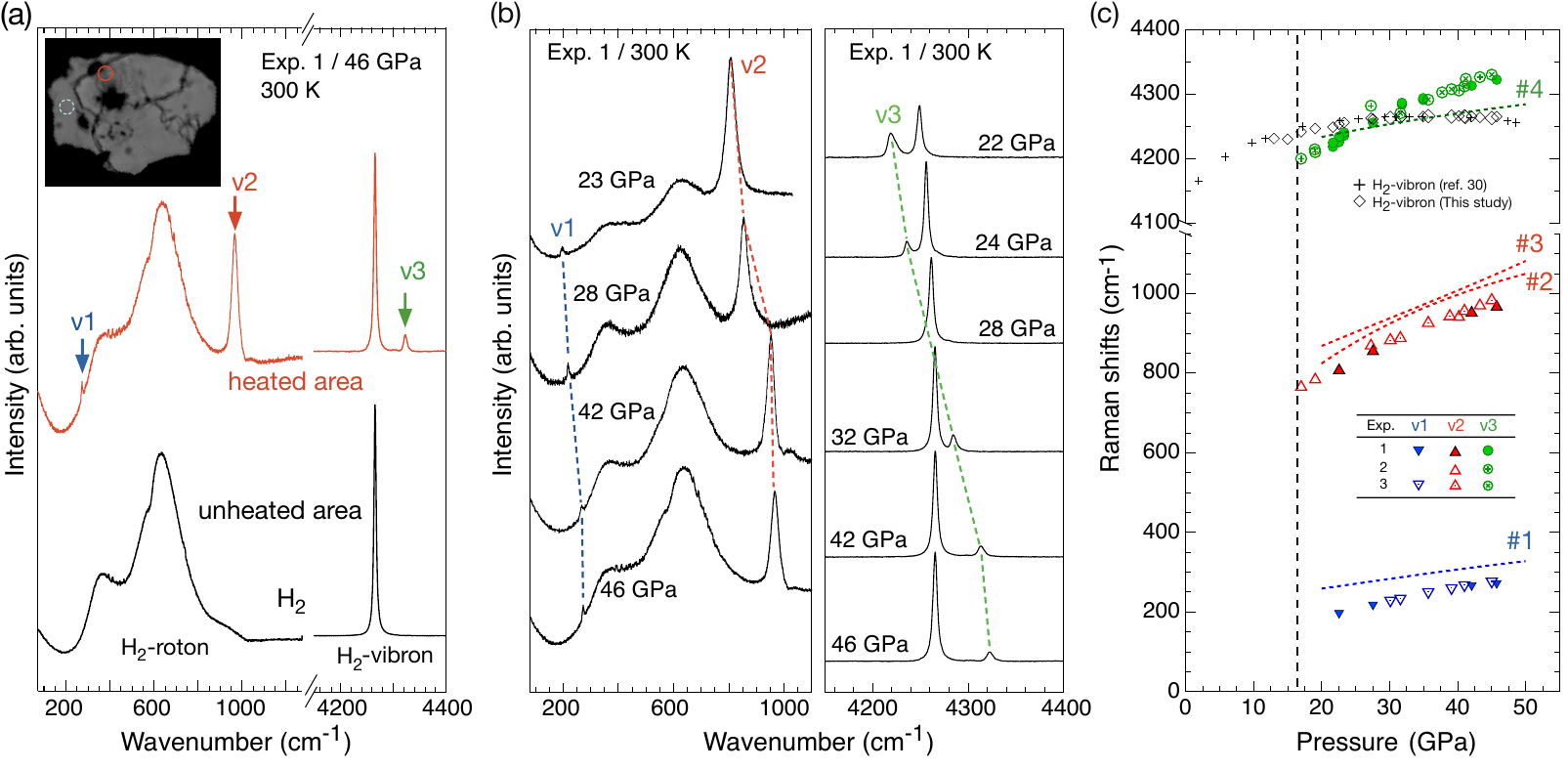}%
  \caption{Raman scattering measurements of NaClH$_{\textit{x}}$ at different pressures. (a) Typical Raman  spectra obtained at the heated area (solid circle) and H$_2$ (dotted circle). (b) Raman spectra of NaClH$_{\textit{x}}$ obtained in Exp. 1 at decreasing pressure. (c) Pressure dependence of the Raman shifts for \(\nu\)1-3 obtained in Exp. 1-3, and H$_2$ vibron of solid H$_2$. The vertical dashed line indicates that NaClH$_{\textit{x}}$ is unstable at the pressures below 17 GPa, and the dotted lines represent the Raman active phonon modes predicted by our first-principles calculations: \(\sharp\)1 (\textit{E}$_{2g}$), \(\sharp\)2 (\textit{E}$_{1g}$), \(\sharp\)3 (\textit{E}$_{2g}$), and \(\sharp\)4 (\textit{A}$_{1g}$).}
  \label{fgr:naclh2_6}
\end{figure}

\subsection{Predicted stable phases of NaCl-hydrogen compounds: NaCl(H$_2$) and NaCl(H$_2$)$_4$ above 20 GPa}
Figure 5a illustrates the formation enthalpy of NaCl-hydrogen compound ((NaCl)$_{1-\textit{x}}$H$_\textit{x}$) from NaCl and H$_2$.
The convex hull diagram indicates that NaCl(H$_2$) emerges as the thermodynamically stable phase above 15 GPa.
Our calculations predict that NaCl(H$_2$) crystallizes in a hexagonal \textit{P}6$_3$/\textit{mmc} structure at 30 GPa with lattice parameters \textit{a} = 3.515 \AA{} and \textit{c} = 6.465 \AA. 
The atomic positions are Na:2\textit{c}(1/3, 2/3, 1/4), Cl:2\textit{a}(0, 0, 0), and  H:4\textit{f}(1/3, 2/3, 0.808).
The Cl and Na layers stack in the ABAC manner along the \textit{c} axis, and H$_2$ molecules occupy the center of mass in the Na triangles in the B (C) layer (Figure 5b).
The predicted crystal symmetry agrees with that of the experimentally observed NaClH$_\textit{x}$.
Besides, the \textit{V}$_{f.u.}$ of NaCl(H$_2$) calculated from the predicted lattice parameters shows good agreement with the experimentally observed NaClH$_\textit{x}$ (Figure 3c).
The volume reduction caused by the formation of NaCl(H$_2$) is 6.1\% at 20 GPa, 1.8\% at 40 GPa, which is in agreement with the experiment, 
and 1.8\% at 50 GPa.

In the crystal structure search, we also identified another hexagonal structure of NaCl(H$_2$) whose space group is  \textit{P}6\textit{m}2 with lattice parameters \textit{a} = 3.507 \AA{} and \textit{c} = 3.250 \AA.
In the \textit{P}6\textit{m}2 structure, the atoms are at Na:1\textit{f}(2/3, 1/3, 1/2), Cl:1\textit{a}(0, 0, 0), and  H:2\textit{h}(1/3, 2/3, 0.615) positions, and the layers of Na and Cl stack in AB manner  (Figure 5c).
The NaCl(H$_2$) with AB stacking possesses a higher enthalpy by 0.15 mRy/atom than that of the ABAC stacking.
Besides, we predicted that a hydrogen-richer phase, NaCl(H$_2$)$_4$, is stabilized at pressures above 40 GPa (Figure 5a and 5d). 
This NaCl(H$_2$)$_4$ is obtained by deforming NaCl(H$_2$) with AB stacking. 
An NaCl(H$_2$) layer is formed by overlapping the B layer with A, and the hexagonal lattice is slightly distorted owing to the H$_2$ molecular axis orientation into the layer. 
Then, H$_2$ molecules are intercalated between the layers, and an (H$_2$)$_3$ layer is created (Figure 5d). 
The space group of this structure is a monoclinic \textit{Pm} with lattice parameters \textit{a} = 4.155 \AA, \textit{b} = 3.146 \AA,  \textit{c} = 4.197 \AA, \(\beta\) = 118.91$^{\circ}$.
The detailed structural parameters for NaCl(H$_2$)$_4$ is shown in the Supporting Information.
Although our calculations predicted the formation of these two phases (\textit{P}6\textit{m}2 and \textit{Pm}), we did not observe them in our experiments.
We will later discuss the difference between experimental and theoretical results in relation to the criteria for the NaCl-hydrogen compound formation.

\begin{figure}
 \includegraphics[scale=1]{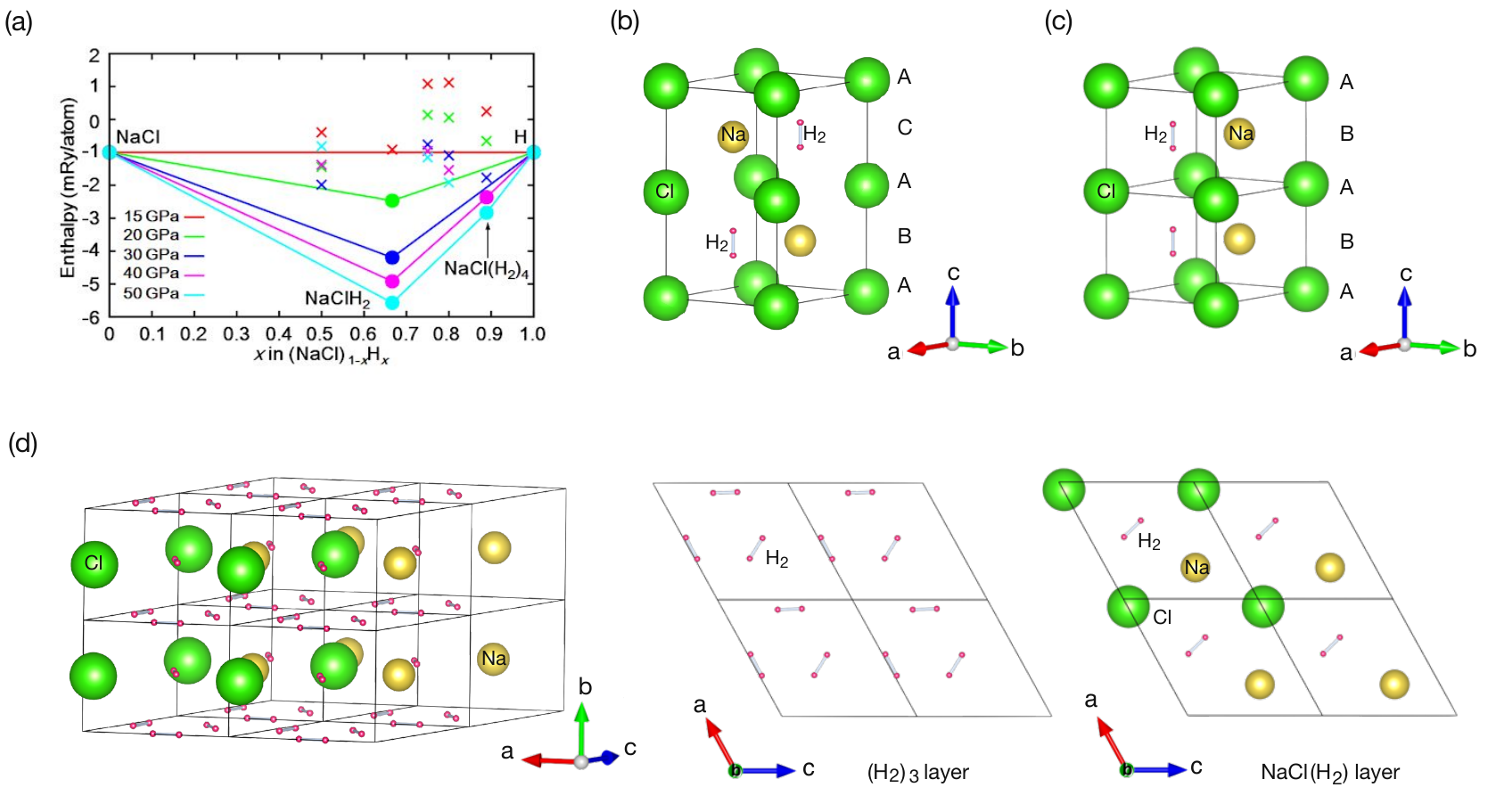}%
  \caption{Crystal structure search by a genetic algorithm technique and first-principles calculations. (a) Formation enthalpy per atom as a function of \textit{x} for (NaCl)$_{1-\textit{x}}$H$_\textit{x}$ at various pressures.  The compounds whose enthalpies (cross marks) are above the convex hull are predicted to decompose. (b) The most stable structure of NaCl(H$_2$), formed by ABAC stacking of a hexagonal layer. Na and Cl are depicted as a cation an anion, respectively. (c) Metastable structure of NaCl(H$_2$), formed by AB stacking. (d) NaCl(H$_2$)$_4$, formed by the stacking of the (H$_2$)$_3$ layer and the NaCl(H$_2$) layer. The crystal structures were drawn using VESTA.\cite{Momma2011}}
  \label{fgr:naclh2_2}
\end{figure}

Figure 6 shows the calculated phonon dispersion of NaCl(H$_2$) with ABAC stacking at different pressures.
Imaginary phonon frequency appears in \(\Gamma\)-M, M-K, K-\(\Gamma\), \(\Gamma\)-A, and A-M lines at 10 GPa (Figure 6a), which indicates that this structure is unstable at this pressure. 
The imaginary phonon frequency disappears above 15 GPa (Figure 6b-d). 
This observation is in agreement with our experiments that the synthesized NaClH$_\textit{x}$ decomposes into NaCl and H$_2$ below 17 GPa.
The phonon at around 4250 cm$^{-1}$ is the H$_2$ vibron that occupies the 4\textit{f} site in NaCl(H$_2$).
This H$_2$ vibron is slightly dispersive at 50 GPa owing to the increases of the interaction between H$_2$ molecules and the other atoms with compression (Figure 6d). 
We note that our first-principles calculations predict that the H$_2$ molecules in NaCl(H$_2$) with ABAC stacking rotationally oscillate with limited angles (librons), and their molecular axes are roughly aligned along the \textit{c} axis.

\begin{figure}
 \includegraphics[scale=1]{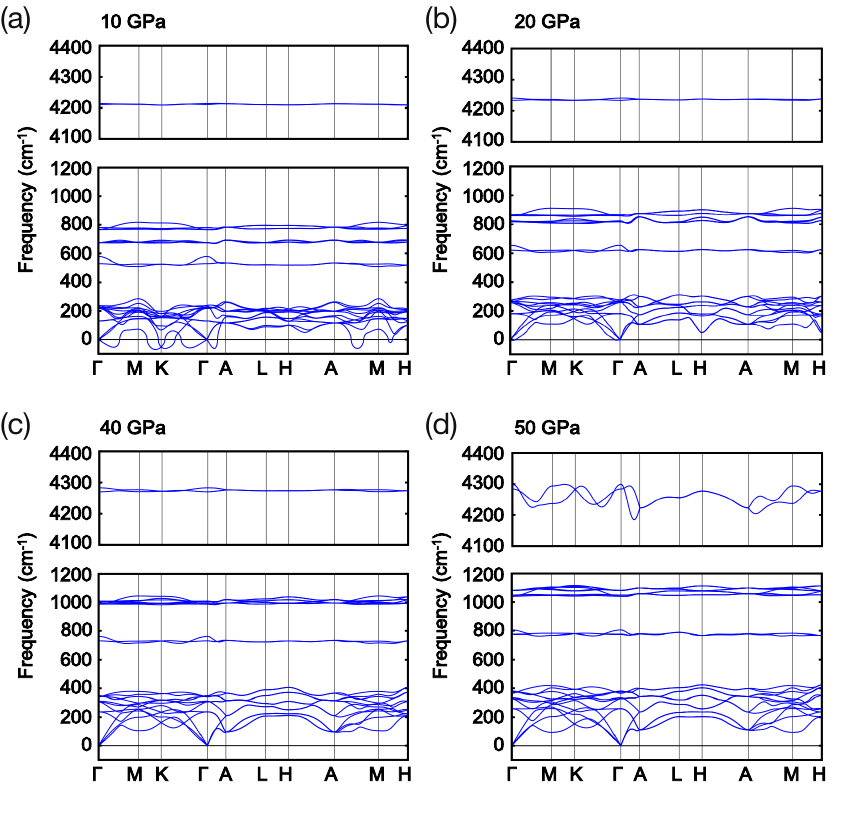}%
  \caption{Phonon dispersion of NaCl(H$_2$) with ABAC stacking at pressures of  (a) 10 GPa, (b) 20 GPa, (c) 40 GPa, and (d) 50 GPa.}
  \label{fgr:naclh2_3}
\end{figure}

In Figure 4c, we also display the predicted Raman-active zone-center phonon modes of NaCl(H$_2$) with ABAC stacking at pressures.
See the Supporting Information for the other phonon modes that are infrared active and silent.
Mode \(\sharp\)1 (\textit{E}$_{2g}$) is the translational oscillation of the B (C) layers in the \textit{a}-\textit{b} plane, and mode  \(\sharp\)2 (\textit{E}$_{1g}$) is the H$_2$ libron. 
Mode  \(\sharp\)3 originates from the translational oscillation of H$_2$ in the \textit{a}-\textit{b} plane, and mode  \(\sharp\)4 is the H$_2$ vibron. 
Here, we compare the prediction of phonon modes  with experimental results.
The predicted frequencies in modes  \(\sharp\)1,  \(\sharp\)2 and \(\sharp\)3, and  \(\sharp\)4 emerge at frequencies similar to  \(\nu\)1,  \(\nu\)2, and  \(\nu\)3, respectively. 
In the Raman spectra,  \(\nu\)2 has a wider peak width than  \(\nu\)1, suggesting the presence of multiple peaks in a broad single peak. 
Assuming that modes  \(\sharp\)2 and \(\sharp\)3 are broadened and overlap,  \(\nu\)2 can be assigned to the overlapped  \(\sharp\)2 and \(\sharp\)3 modes. 
In addition, the predicted positive pressure dependence of modes  \(\sharp\)1-3 agrees with that of  \(\nu\)1-3 (Figure 4c). 
We also note that the pressure dependence of the observed H$_2$ vibron shows an excellent agreement with the reported data of pure H$_2$\cite{Mao1994a}, which supports our conclusion that the observed Raman peaks except for \(\nu\)1-3 are from solid H$_2$.

\section{DISCUSSION}
Considering all experimental and theoretical data together, we conclude that hexagonal NaCl(H$_2$) with ABAC stacking is synthesized by heating the NaCl/H$_2$ mixture to approximately 1500 K above 40 GPa. 
While a pressure near 40 GPa is necessary for synthesis, NaCl(H$_2$) is stable down to 17 GPa. 

The questions remained are; (\textit{i}) What kinds of the chemical bonding are formed between H$_2$ and NaCl host lattice?, (\textit{ii}) How the compound is formed at high pressure and high temperatures?, (\textit{iii}) What is the criteria for NaCl-hydrogen compound formation?, and (\textit{iv}) Why the salt is incompatible with hydrogen at room temperature?

On the question (i), the Raman scattering measurements show that the H$_2$-vibron frequency of NaCl(H$_2$) continues to increase beyond that of pure solid hydrogen with the increase of pressure, which implies that there is little interaction between NaCl host lattice and H$_2$ molecules. 
The results of the phonon calculations also support this hypothesis, in which the H$_2$-vibron frequency is non-dispersive up to at least 40 GPa. 
That is to say, NaCl host lattice acts as a compressor for the H$_2$ molecule up to at least 46 GPa, similarly to the case of argon-hydrogen compounds reported earlier.\cite{Ishikawa2017a}

On the question (ii), we speculate that the NaCl(H$_2$) is formed by the insertion of H$_2$ molecules into the B2-type NaCl host lattice.
The temperature used in the present study was about 1500 K which is below the melting point of NaCl at 42-46 GPa\cite{Boehler1997} and well above the melting point of H$_2$.\cite{Dzyabura2013}
Therefore, NaCl(H$_2$) is a product of the reaction between solid B2-type NaCl and fluid H$_2$.
The \textit{P}6$_3$/\textit{mmc} structure for NaCl(H$_2$) can be obtained by the small distortion of the B2-type structure and the displacement of the body-centered Na atom.

To obtain additional insight relating to the question (iii), we laser-heated (2000 K) cesium chloride (CsCl) at 7 GPa and potassium chloride (KCl) at 17 GPa.
Both of these compounds have a B2 structure. 
KCl at 17 GPa has a larger interstitial site than NaCl at 46 GPa. 
In contrast, CsCl has inadequate interstitial site volume to accommodate an H$_2$ molecule.
KCl appears to have the potential to form a KCl(H$_2$)$_\textit{x}$ while CsCl does not.
However, no structural transformations were observed for both KCl and CsCl. 
Although further studies are necessary, we think that many factors such as crystal structure, the ionic radii of elements that form salts, and pressure and temperatures dictate the formation criteria of salt-hydrogen compounds.

Answering the question (iv) is still difficult at current. 
The laser heating at 25 GPa, where NaCl is in B2-type structure at temperatures above 1100 K,\cite{Nishiyama2003} did not result in the formation of the NaClH$_\textit{x}$, while our theoretical calculations predicted NaCl(H$_2$) becomes stable above 15 GPa.
Along with the fact that the formation of NaClH$_\textit{x}$ is not induced by the compression  of the NaCl/H$_2$ mixture at room temperature,  it is thought that there is a large energy barrier between NaCl+H$_2$ and NaClH$_\textit{x}$.
Therefore, sufficiently high temperature is necessary to overcome the barrier depending on pressure.
It would be worth trying the synthesis of NaCl(H$_2$) with AB stacking and NaCl(H$_2$)$_4$, which are predicted by first-principles calculations, at much higher pressures and temperatures.
The information of the present study along with the studies of other ionic compounds\cite{Matsuoka2017, Struzhkin2016} will contribute to a deeper understanding of the inertness against H$_2$.

\section{CONCLUSION}
The present study showed that NaCl can indeed form a compound with hydrogen, NaClH$_\textit{x}$.
The NaClH$_\textit{x}$ is suggested to be NaCl(H$_2$) which has a \textit{P}6$_3$/\textit{mmc} structure accommodating two H$_2$ molecules in the interstitial sites of the unit cell.
Upon the decrease of pressure at 300 K, NaCl(H$_{2}$) remains stable at pressures down to 17 GPa.
Such a large hysteresis in pressure at room temperature suggests the possibility of recovering NaCl(H$_2$) to ambient pressure at low temperatures when the mobility of hydrogen is significantly suppressed.
Our calculations predict that hydrogen-richer phase NaCl(H$_2$)$_4$ is stabilized at pressures above 40 GPa.
A further question remains whether the synthesized NaClH$_\textit{x}$ phase has the stoichiometric composition NaCl(H$_2$). 
\textit{In-situ} neutron diffraction measurements, which include the recovery of materials at low temperature, are expected to provide an answer to the question.


\providecommand{\latin}[1]{#1}
\makeatletter
\providecommand{\doi}
  {\begingroup\let\do\@makeother\dospecials
  \catcode`\{=1 \catcode`\}=2 \doi@aux}
\providecommand{\doi@aux}[1]{\endgroup\texttt{#1}}
\makeatother
\providecommand*\mcitethebibliography{\thebibliography}
\csname @ifundefined\endcsname{endmcitethebibliography}
  {\let\endmcitethebibliography\endthebibliography}{}

\newpage
\section{Supporting Information - Hydrogen-Storing Salt NaCl(H$_2$) Synthesized at High Pressure and High Temperature}

\setcounter{page}{1}
\renewcommand{\thepage}{\Alph{section}\arabic{page}}


\setcounter{section}{19}
\renewcommand{\thesection}{\Alph{section}}
\setcounter{figure}{0} 
\renewcommand{\thefigure}{\Alph{section}\arabic{figure}}
\setcounter{table}{0}
\renewcommand{\thetable}{\Alph{section}\arabic{table}}
\renewcommand*{\citenumfont}[1]{S#1}
\renewcommand*{\bibnumfmt}[1]{(S#1)}

\tableofcontents
\begin{itemize}
  \item Structural parameters for NaCl(H$_2$) and NaCl(H$_2$)$_4$ -results of \textit{ab-initio} calculations-
  \item Comparison of Raman scattering data between NaCl(H$_\textit{x}$), Na$_3$Cl, NaCl$_3$, and NaH$_\textit{x}$ (\textit{x} = 3, 7)
  \item Calculated zone-center phonon frequencies of NaCl(H$_2$) with ABAC-stacking at 46 GPa
  \end{itemize}

\newpage 
  \subsection{Structural parameters for NaCl(H$_2$) and NaCl(H$_2$)$_4$ -results of \textit{ab-initio} calculations-}
  
  \begin{table}
  \caption{NaCl(H$_2$) with ABAC stacking at 30 GPa}
  \label{tbl:example}
  \begin{tabular}{lclll}
    \hline
     \multicolumn{2}{l}{Space group \textit{P}6$_3$/\textit{mmc},} & \multicolumn{3}{l}{\textit{a} = 3.515 \AA, \textit{c} = 6.465 \AA} \\
         & & \textit{x} & \textit{y} & \textit{z} \\
    Na   & 2\textit{c}  & 1/3 & 2/3 & 1/4\\
    Cl & 2\textit{a}  & 0 & 0 & 0\\
    H  & 4\textit{f}   & 1/3 & 2/3 & 0.808\\
    \hline
  \end{tabular}
\end{table}

  \begin{table}
  \caption{NaCl(H$_2$) with AB stacking at 30 GPa}
  \label{tbl:example}
  \begin{tabular}{lclll}
    \hline
    \multicolumn{2}{l}{Space group \textit{P}6\textit{m}2,} & \multicolumn{3}{l}{\textit{a} = 3.507 \AA, \textit{c} = 3.250 \AA} \\
         & &  \textit{x} & \textit{y} & \textit{z} \\
    Na   & 1\textit{f}  & 2/3 & 1/3 & 1/2 \\
    Cl & 1\textit{a}  & 0 & 0 & 0 \\
    H  & 2\textit{h}   & 1/3 & 2/3 & 0.615\\
    \hline
  \end{tabular}
\end{table}

  \begin{table}
  \caption{NaCl(H$_2$)$_4$ at 50 GPa}
  \label{tbl:example}
  \begin{tabular}{lclll}
    \hline
    \multicolumn{2}{l}{Space group \textit{Pm},} & \multicolumn{3}{l}{\textit{a} = 4.155 \AA, \textit{b} = 3.146 \AA}\\
    && \multicolumn{3}{l}{\textit{c} = 4.197 \AA, $\beta$ = 118.91$^{\circ}$}\\
         &&   \textit{x} & \textit{y} & \textit{z} \\
    Na   & 1\textit{b} &  0.26767 & 1/2 & 0.66311\\
    Cl & 1\textit{b}   & -0.06614 & 1/2 & 0.00651\\
    H & 1\textit{a} & -0.08452 & 0 & 0.40644\\
    H & 1\textit{a} & -0.07692 & 0 & 0.58805\\
    H & 1\textit{b} & 0.67073 & 1/2 & 0.45016\\
    H & 1\textit{b} & 0.53023 & 1/2 & 0.25898\\
    H & 1\textit{a} & 0.53226 & 0 & 0.02106\\
    H & 1\textit{a} & 0.52772 & 0 & 0.58598\\
    H & 1\textit{a} & 0.35007 & 0 & 0.01730\\
    H & 1\textit{a} & 0.35405 & 0 & 0.40858\\
    \hline
  \end{tabular}
\end{table}
  
\newpage  
  \subsection{Comparison of Raman scattering data between NaCl(H$_\textit{x}$), Na$_3$Cl, NaCl$_3$, and NaH$_\textit{x}$ (\textit{x} = 3, 7)}
  
  Here, we compare the Raman scattering spectra obtained for  NaCl(H$_\textit{x}$) with the data of Na$_3$Cl, NaCl$_3$, and NaH$_\textit{x}$ (\textit{x} = 3, 7). The Raman scattering spectrum and the pressure dependence of the observed Raman scattering peaks of NaCl(H$_\textit{x}$) do not match the data for any of Na$_3$Cl, NaCl$_3$, and NaH$_\textit{x}$ (\textit{x} = 3, 7).
\begin{figure}
 \includegraphics[scale=1]{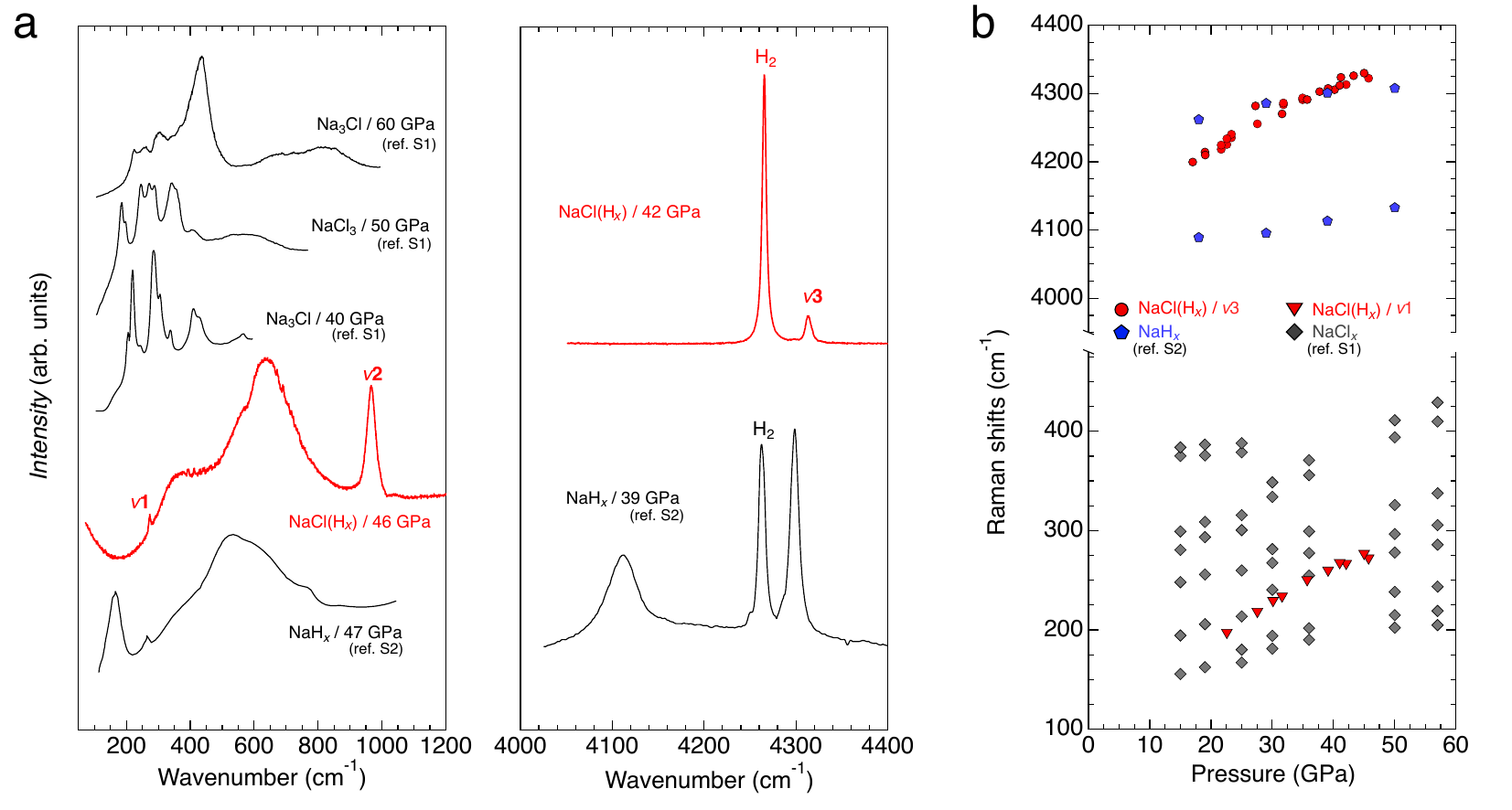}%
  \caption{(a) Raman scattering spectra of NaCl(H$_\textit{x}$), Na$_3$Cl\cite{Zhang}, NaCl$_3$\cite{Zhang}, and NaH$_\textit{x}$ (\textit{x} = 3, 7).\cite{Struzhkin2016}  The broad peak, which is seen in the spectra of NaCl(H$_\textit{x}$) and NaH$_\textit{x}$, centering at 600 cm$^{-1}$ is from the roton of excess pure H$_2$. The sharp peak centering at 4260 cm$^{-1}$ is the vibron of excess pure H$_2$. The data of Na$_3$Cl, NaCl$_3$, and NaH$_\textit{x}$ were taken from refs. S1 and S2, and their supplementary materials. (b) Pressure dependence of the Raman shifts of \(\nu\)1 and \(\nu\)3 plotted together with the data of NaCl$_\textit{x}$ and NaH$_\textit{x}$. The data of NaCl$_\textit{x}$ has been assigned to Na$_3$Cl in Figure S10 of ref. S1. The data of NaH$_\textit{x}$ has been reported to contain the contributions from both NaH$_3$ and NaH$_7$.\cite{Struzhkin2016}}
  \label{fgr:naclhx_S1}
\end{figure}

\newpage
  
  \subsection{Calculated zone-center phonon frequencies of NaCl(H$_2$) with ABAC-stacking at 46 GPa}
   
  \begin{table}
  \caption{Raman active modes}
  \label{tbl:example}
  \begin{tabular}{lclll}
    \hline
    Mode & \(\omega\)$_{calc.}$ (cm$^{-1}$) \\
    \hline
    \textit{E}$_{2g}$  & 319.7  \\
    \textit{E}$_{1g}$  & 1030.0  \\
    \textit{E}$_{2g}$   & 1053.7  \\
    \textit{A}$_{1g}$  & 4278.7  \\
    \hline
  \end{tabular}
\end{table}
  
  \begin{table}
  \caption{Infrared active modes}
  \label{tbl:example}
  \begin{tabular}{lclll}
   \hline
    Mode & \(\omega\)$_{calc.}$ (cm$^{-1}$)\\
    \hline
    \textit{A}$_{2u}$  & 74.8  \\
    \textit{E}$_{1u}$  & 97.9  \\
    \textit{A}$_{2u}$   & 361.1  \\
    \textit{E}$_{1u}$  & 369.5  \\
    \textit{A}$_{2u}$   & 751.2  \\
    \textit{E}$_{1u}$  & 1047.9  \\
    \hline
  \end{tabular}
\end{table}

\begin{table}
  \caption{Silent modes}
  \label{tbl:example}
  \begin{tabular}{lclll}
   \hline
    Mode & \(\omega\)$_{calc.}$ (cm$^{-1}$)\\
    \hline
    \textit{E}$_{2u}$  & 251.0  \\
    \textit{B}$_{2g}$  & 326.6  \\
    \textit{B}$_{1u}$   & 354.9  \\
    \textit{B}$_{2g}$  & 760.2  \\
    \textit{E}$_{2u}$  & 1021.0  \\
    \textit{B}$_{1u}$  & 4293.1  \\
    \hline
  \end{tabular}
\end{table}
 

\begin{mcitethebibliography}{33}
\providecommand*\natexlab[1]{#1}
\providecommand*\mciteSetBstSublistMode[1]{}
\providecommand*\mciteSetBstMaxWidthForm[2]{}
\providecommand*\mciteBstWouldAddEndPuncttrue
  {\def\EndOfBibitem{\unskip.}}
\providecommand*\mciteBstWouldAddEndPunctfalse
  {\let\EndOfBibitem\relax}
\providecommand*\mciteSetBstMidEndSepPunct[3]{}
\providecommand*\mciteSetBstSublistLabelBeginEnd[3]{}
\providecommand*\EndOfBibitem{}
\mciteSetBstSublistMode{f}
\mciteSetBstMaxWidthForm{subitem}{(\alph{mcitesubitemcount})}
\mciteSetBstSublistLabelBeginEnd
  {\mcitemaxwidthsubitemform\space}
  {\relax}
  {\relax}

\bibitem[Feng \latin{et~al.}(2006)Feng, Grochala, Jaro{\'{n}}, Hoffmann,
  Bergara, and Ashcroft]{Feng2006}
Feng,~J.; Grochala,~W.; Jaro{\'{n}},~T.; Hoffmann,~R.; Bergara,~A.;
  Ashcroft,~N.~W. {Structures and Potential Superconductivity in SiH$_4$ at
  High Pressure: En Route to ''Metallic Hydrogen''}. \emph{Phys. Rev.
  Lett.} \textbf{2006}, \emph{96}, 017006--1--4\relax
\mciteBstWouldAddEndPuncttrue
\mciteSetBstMidEndSepPunct{\mcitedefaultmidpunct}
{\mcitedefaultendpunct}{\mcitedefaultseppunct}\relax
\EndOfBibitem
\bibitem[Peng \latin{et~al.}(2017)Peng, Sun, Pickard, Needs, Wu, and
  Ma]{Peng2017}
Peng,~F.; Sun,~Y.; Pickard,~C.~J.; Needs,~R.~J.; Wu,~Q.; Ma,~Y. {Hydrogen
  Clathrate Structures in Rare Earth Hydrides at High Pressures: Possible Route
  to Room-Temperature Superconductivity}. \emph{Phys. Rev. Lett.}
  \textbf{2017}, \emph{119}, 107001--1--6\relax
\mciteBstWouldAddEndPuncttrue
\mciteSetBstMidEndSepPunct{\mcitedefaultmidpunct}
{\mcitedefaultendpunct}{\mcitedefaultseppunct}\relax
\EndOfBibitem
\bibitem[Li \latin{et~al.}(2014)Li, Hao, Liu, Li, and Ma]{Li2014}
Li,~Y.; Hao,~J.; Liu,~H.; Li,~Y.; Ma,~Y. {The Metallization and
  Superconductivity of Dense Hydrogen Sulfide.} \emph{J. Chem. Phys.} \textbf{2014}, \emph{140}, 174712--1--7\relax
\mciteBstWouldAddEndPuncttrue
\mciteSetBstMidEndSepPunct{\mcitedefaultmidpunct}
{\mcitedefaultendpunct}{\mcitedefaultseppunct}\relax
\EndOfBibitem
\bibitem[Drozdov \latin{et~al.}(2015)Drozdov, Eremets, Troyan, Ksenofontov, and
  Shylin]{Drozdov2015}
Drozdov,~A.~P.; Eremets,~M.~I.; Troyan,~I.~A.; Ksenofontov,~V.; Shylin,~S.~I.
  {Conventional Superconductivity at 203 Kelvin at High Pressures in the Sulfur
  Hydride System}. \emph{Nature} \textbf{2015}, \emph{525}, 73--76\relax
\mciteBstWouldAddEndPuncttrue
\mciteSetBstMidEndSepPunct{\mcitedefaultmidpunct}
{\mcitedefaultendpunct}{\mcitedefaultseppunct}\relax
\EndOfBibitem
\bibitem[Somayazulu \latin{et~al.}(2018)Somayazulu, Ahart, Mishra, Geballe,
  Baldini, Meng, Struzhkin, and Hemley]{Somayazulu2018a}
Somayazulu,~M.; Ahart,~M.; Mishra,~A.~K.; Geballe,~Z.~M.; Baldini,~M.;
  Meng,~Y.; Struzhkin,~V.~V.; Hemley,~R.~J. {Evidence for Superconductivity
  above 260 K in Lanthanum Superhydride at Megabar Pressures}. \emph{Phys.
  Rev. Lett.} \textbf{2018}, \emph{122}, 27001--1--6\relax
\mciteBstWouldAddEndPuncttrue
\mciteSetBstMidEndSepPunct{\mcitedefaultmidpunct}
{\mcitedefaultendpunct}{\mcitedefaultseppunct}\relax
\EndOfBibitem
\bibitem[Drozdov \latin{et~al.}(2019)Drozdov, Kong, Minkov, Besedin,
  Kuzovnikov, Mozaffari, Balicas, Balakirev, Graf, Prakapenka, Greenberg,
  Knyazev, Tkacz, and Eremets]{Drozdov2018a}
Drozdov,~A.~P.; Kong,~P.~P.; Minkov,~V.~S.; Besedin,~S.~P.; Kuzovnikov,~M.~A.;
  Mozaffari,~S.; Balicas,~L.; Balakirev,~F.~F.; Graf,~D.~E.; Prakapenka,~V.~B.
  \latin{et~al.}  {Superconductivity at 250 K in Lanthanum Hydride under High
  Pressures}. \emph{Nature} \textbf{2019}, \emph{569}, 528--531\relax
\mciteBstWouldAddEndPuncttrue
\mciteSetBstMidEndSepPunct{\mcitedefaultmidpunct}
{\mcitedefaultendpunct}{\mcitedefaultseppunct}\relax
\EndOfBibitem
\bibitem[Ohta \latin{et~al.}(2015)Ohta, Ichimaru, Einaga, Kawaguchi, Shimizu,
  Matsuoka, Hirao, and Ohishi]{Ohta2015}
Ohta,~K.; Ichimaru,~K.; Einaga,~M.; Kawaguchi,~S.; Shimizu,~K.; Matsuoka,~T.;
  Hirao,~N.; Ohishi,~Y. {Phase Boundary of Hot Dense Fluid Hydrogen}.
  \emph{Sci. Rep.} \textbf{2015}, \emph{5}, 16560--1--7\relax
\mciteBstWouldAddEndPuncttrue
\mciteSetBstMidEndSepPunct{\mcitedefaultmidpunct}
{\mcitedefaultendpunct}{\mcitedefaultseppunct}\relax
\EndOfBibitem
\bibitem[Sakamaki \latin{et~al.}(2009)Sakamaki, Takahashi, Nakajima, Nishihara,
  Funakoshi, Suzuki, and Fukai]{Sakamaki2009}
Sakamaki,~K.; Takahashi,~E.; Nakajima,~Y.; Nishihara,~Y.; Funakoshi,~K.;
  Suzuki,~T.; Fukai,~Y. {Melting Phase Relation of FeH$_\textit{x}$ up to
  20GPa: Implication for the Temperature of the Earth's Core}. \emph{Phys. Earth Planet. Inter.} \textbf{2009}, \emph{174}, 192--201\relax
\mciteBstWouldAddEndPuncttrue
\mciteSetBstMidEndSepPunct{\mcitedefaultmidpunct}
{\mcitedefaultendpunct}{\mcitedefaultseppunct}\relax
\EndOfBibitem
\bibitem[Donnerer \latin{et~al.}(2013)Donnerer, Scheler, and
  Gregoryanz]{Donnerer2013}
Donnerer,~C.; Scheler,~T.; Gregoryanz,~E. {High-Pressure Synthesis of Noble
  Metal Hydrides}. \emph{J. Chem. Phys.} \textbf{2013},
  \emph{138}, 134507--1--6\relax
\mciteBstWouldAddEndPuncttrue
\mciteSetBstMidEndSepPunct{\mcitedefaultmidpunct}
{\mcitedefaultendpunct}{\mcitedefaultseppunct}\relax
\EndOfBibitem
\bibitem[Matsuoka \latin{et~al.}(2011)Matsuoka, Hirao, Ohishi, Shimizu,
  Machida, and Aoki]{Matsuoka2011c}
Matsuoka,~T.; Hirao,~N.; Ohishi,~Y.; Shimizu,~K.; Machida,~A.; Aoki,~K.
  {Structural and Electrical Transport Properties of FeH$_x$ under High
  Pressures and Low Temperatures}. \emph{High Pressure Res.} \textbf{2011},
  \emph{31}, 64--67\relax
\mciteBstWouldAddEndPuncttrue
\mciteSetBstMidEndSepPunct{\mcitedefaultmidpunct}
{\mcitedefaultendpunct}{\mcitedefaultseppunct}\relax
\EndOfBibitem
\bibitem[Pauling(1960)]{Pauling1960}
Pauling,~L. \emph{The Nature of the Chemical Bond and the Structure of
  Molecules and Crystals : An Introduction to Modern Structural Chemistry}, 3rd
  ed.; Cornell University Press, USA, 1960\relax
\mciteBstWouldAddEndPuncttrue
\mciteSetBstMidEndSepPunct{\mcitedefaultmidpunct}
{\mcitedefaultendpunct}{\mcitedefaultseppunct}\relax
\EndOfBibitem
\bibitem[Kuno \latin{et~al.}(2015)Kuno, Matsuoka, Nakagawa, Hirao, Ohishi,
  Shimizu, Takahama, Ohta, Sakata, Nakamoto, Kume, and Sasaki]{Kuno2015}
Kuno,~K.; Matsuoka,~T.; Nakagawa,~T.; Hirao,~N.; Ohishi,~Y.; Shimizu,~K.;
  Takahama,~K.; Ohta,~K.; Sakata,~M.; Nakamoto,~Y. \latin{et~al.}  {Heating of
  Li in Hydrogen: Possible Synthesis of LiH$_\textit{x}$}. \emph{High Pressure Res.} \textbf{2015}, \emph{35}, 16--21\relax
\mciteBstWouldAddEndPuncttrue
\mciteSetBstMidEndSepPunct{\mcitedefaultmidpunct}
{\mcitedefaultendpunct}{\mcitedefaultseppunct}\relax
\EndOfBibitem
\bibitem[Matsuoka \latin{et~al.}(2017)Matsuoka, Kuno, Ohta, Sakata, Nakamoto,
  Hirao, Ohishi, Shimizu, Kume, and Sasaki]{Matsuoka2017}
Matsuoka,~T.; Kuno,~K.; Ohta,~K.; Sakata,~M.; Nakamoto,~Y.; Hirao,~N.;
  Ohishi,~Y.; Shimizu,~K.; Kume,~T.; Sasaki,~S. {Lithium Polyhydrides
  Synthesized under High Pressure and High Temperature}. \emph{J. Raman Spectrosc.} \textbf{2017}, \emph{48}, 1222--1228\relax
\mciteBstWouldAddEndPuncttrue
\mciteSetBstMidEndSepPunct{\mcitedefaultmidpunct}
{\mcitedefaultendpunct}{\mcitedefaultseppunct}\relax
\EndOfBibitem
\bibitem[Struzhkin \latin{et~al.}(2016)Struzhkin, Kim, Stavrou, Muramatsu, Mao,
  Pickard, Needs, Prakapenka, and Goncharov]{Struzhkin2016}
Struzhkin,~V.~V.; Kim,~D.~Y.; Stavrou,~E.; Muramatsu,~T.; Mao,~H.-k.;
  Pickard,~C.~J.; Needs,~R.~J.; Prakapenka,~V.~B.; Goncharov,~A.~F. {Synthesis
  of Sodium Polyhydrides at High Pressures}. \emph{Nat. Commun.}
  \textbf{2016}, \emph{7}, 12267--1--8\relax
\mciteBstWouldAddEndPuncttrue
\mciteSetBstMidEndSepPunct{\mcitedefaultmidpunct}
{\mcitedefaultendpunct}{\mcitedefaultseppunct}\relax
\EndOfBibitem
\bibitem[Chi \latin{et~al.}(2011)Chi, Nguyen, Matsuoka, Kagayama, Hirao,
  Ohishi, and Shimizu]{Chi2011}
Chi,~Z.; Nguyen,~H.; Matsuoka,~T.; Kagayama,~T.; Hirao,~N.; Ohishi,~Y.;
  Shimizu,~K. {Cryogenic Implementation of Charging Diamond Anvil Cells with
  H$_2$ and D$_2$.} \emph{Rev. Sci. Instrum.} \textbf{2011},
  \emph{82}, 105109--1--4\relax
\mciteBstWouldAddEndPuncttrue
\mciteSetBstMidEndSepPunct{\mcitedefaultmidpunct}
{\mcitedefaultendpunct}{\mcitedefaultseppunct}\relax
\EndOfBibitem
\bibitem[Ohishi \latin{et~al.}(2008)Ohishi, Hirao, Sata, Hirose, and
  Takata]{Ohishi2008}
Ohishi,~Y.; Hirao,~N.; Sata,~N.; Hirose,~K.; Takata,~M. {Highly Intense
  Monochromatic X-ray Diffraction Facility for High-Pressure Research at
  SPring-8}. \emph{High Pressure Res.} \textbf{2008}, \emph{28},
  163--173\relax
\mciteBstWouldAddEndPuncttrue
\mciteSetBstMidEndSepPunct{\mcitedefaultmidpunct}
{\mcitedefaultendpunct}{\mcitedefaultseppunct}\relax
\EndOfBibitem
\bibitem[Akahama and Kawamura(2006)Akahama, and Kawamura]{Akahama2006}
Akahama,~Y.; Kawamura,~H. {Pressure Calibration of Diamond Anvil Raman Gauge to
  310 GPa}. \emph{J. Appl. Phys.} \textbf{2006}, \emph{100},
  043516--1--4\relax
\mciteBstWouldAddEndPuncttrue
\mciteSetBstMidEndSepPunct{\mcitedefaultmidpunct}
{\mcitedefaultendpunct}{\mcitedefaultseppunct}\relax
\EndOfBibitem
\bibitem[Giannozzi \latin{et~al.}(2017)Giannozzi, Andreussi, Brumme, Bunau,
  {Buongiorno Nardelli}, Calandra, Car, Cavazzoni, Ceresoli, Cococcioni,
  Colonna, Carnimeo, {Dal Corso}, de~Gironcoli, Delugas, DiStasio, Ferretti,
  Floris, Fratesi, Fugallo, Gebauer, Gerstmann, Giustino, Gorni, Jia, Kawamura,
  Ko, Kokalj, K{\"{u}}{\c{c}}{\"{u}}kbenli, Lazzeri, Marsili, Marzari, Mauri,
  Nguyen, Nguyen, Otero-de-la Roza, Paulatto, Ponc{\'{e}}, Rocca, Sabatini,
  Santra, Schlipf, Seitsonen, Smogunov, Timrov, Thonhauser, Umari, Vast, Wu,
  and Baroni]{Giannozzi2017}
Giannozzi,~P.; Andreussi,~O.; Brumme,~T.; Bunau,~O.; {Buongiorno Nardelli},~M.;
  Calandra,~M.; Car,~R.; Cavazzoni,~C.; Ceresoli,~D.; Cococcioni,~M.
  \latin{et~al.}  {Advanced Capabilities for Materials Modelling with Quantum
  ESPRESSO}. \emph{J. Phys.: Condens. Matter} \textbf{2017},
  \emph{29}, 465901--1--30\relax
\mciteBstWouldAddEndPuncttrue
\mciteSetBstMidEndSepPunct{\mcitedefaultmidpunct}
{\mcitedefaultendpunct}{\mcitedefaultseppunct}\relax
\EndOfBibitem
\bibitem[Giannozzi \latin{et~al.}(2009)Giannozzi, Baroni, Bonini, Calandra,
  Car, Cavazzoni, Ceresoli, Chiarotti, Cococcioni, Dabo, {Dal Corso},
  de~Gironcoli, Fabris, Fratesi, Gebauer, Gerstmann, Gougoussis, Kokalj,
  Lazzeri, Martin-Samos, Marzari, Mauri, Mazzarello, Paolini, Pasquarello,
  Paulatto, Sbraccia, Scandolo, Sclauzero, Seitsonen, Smogunov, Umari, and
  Wentzcovitch]{Giannozzi2009}
Giannozzi,~P.; Baroni,~S.; Bonini,~N.; Calandra,~M.; Car,~R.; Cavazzoni,~C.;
  Ceresoli,~D.; Chiarotti,~G.~L.; Cococcioni,~M.; Dabo,~I. \latin{et~al.}
  {QUANTUM ESPRESSO: A Modular and Open-Source Software Project for Quantum
  Simulations of Materials}. \emph{J. Phys.: Condens. Matter}
  \textbf{2009}, \emph{21}, 395502--1--19\relax
\mciteBstWouldAddEndPuncttrue
\mciteSetBstMidEndSepPunct{\mcitedefaultmidpunct}
{\mcitedefaultendpunct}{\mcitedefaultseppunct}\relax
\EndOfBibitem
\bibitem[Ishikawa \latin{et~al.}(2016)Ishikawa, Nakanishi, Shimizu,
  Katayama-Yoshida, Oda, and Suzuki]{Ishikawa2016a}
Ishikawa,~T.; Nakanishi,~A.; Shimizu,~K.; Katayama-Yoshida,~H.; Oda,~T.;
  Suzuki,~N. {Superconducting H$_5$S$_2$ Phase in Sulfur-Hydrogen System under
  High-Pressure}. \emph{Sci. Rep.} \textbf{2016}, \emph{6}, 23160--1--8\relax
\mciteBstWouldAddEndPuncttrue
\mciteSetBstMidEndSepPunct{\mcitedefaultmidpunct}
{\mcitedefaultendpunct}{\mcitedefaultseppunct}\relax
\EndOfBibitem
\bibitem[Ishikawa \latin{et~al.}(2017)Ishikawa, Nakanishi, Shimizu, and
  Oda]{Ishikawa2017a}
Ishikawa,~T.; Nakanishi,~A.; Shimizu,~K.; Oda,~T. {Phase Stability and
  Superconductivity of Compressed Argon-Hydrogen Compounds from
  First-Principles}. \emph{J. Phys. Soc. Jpn.}
  \textbf{2017}, \emph{86}, 124711--1--5\relax
\mciteBstWouldAddEndPuncttrue
\mciteSetBstMidEndSepPunct{\mcitedefaultmidpunct}
{\mcitedefaultendpunct}{\mcitedefaultseppunct}\relax
\EndOfBibitem
\bibitem[Perdew \latin{et~al.}(1996)Perdew, Burke, and Ernzerhof]{Perdew1996}
Perdew,; Burke,; Ernzerhof, {Generalized Gradient Approximation Made Simple.}
  \emph{Phys. Rev. Lett.} \textbf{1996}, \emph{77}, 3865--3868\relax
\mciteBstWouldAddEndPuncttrue
\mciteSetBstMidEndSepPunct{\mcitedefaultmidpunct}
{\mcitedefaultendpunct}{\mcitedefaultseppunct}\relax
\EndOfBibitem
\bibitem[Vanderbilt(1990)]{Vanderbilt1990}
Vanderbilt,~D. {Soft Self-Consistent Pseudopotentials in a Generalized
  Eigenvalue Formalism}. \emph{Phys. Rev. B} \textbf{1990}, \emph{41},
  7892--7895\relax
\mciteBstWouldAddEndPuncttrue
\mciteSetBstMidEndSepPunct{\mcitedefaultmidpunct}
{\mcitedefaultendpunct}{\mcitedefaultseppunct}\relax
\EndOfBibitem
\bibitem[Bassett(1968)]{Bassett1968}
Bassett,~W.~A. {Pressure-Induced Phase Transformation in NaCl}. \emph{J. Appl. Phys.} \textbf{1968}, \emph{39}, 319--325\relax
\mciteBstWouldAddEndPuncttrue
\mciteSetBstMidEndSepPunct{\mcitedefaultmidpunct}
{\mcitedefaultendpunct}{\mcitedefaultseppunct}\relax
\EndOfBibitem
\bibitem[Li and Jeanloz(1987)Li, and Jeanloz]{Li1987}
Li,~X.; Jeanloz,~R. {Measurement of the B1- B2 Transition Pressure in NaCl at
  High Temperatures}. \emph{Phys. Rev. B} \textbf{1987}, \emph{36},
  474--479\relax
\mciteBstWouldAddEndPuncttrue
\mciteSetBstMidEndSepPunct{\mcitedefaultmidpunct}
{\mcitedefaultendpunct}{\mcitedefaultseppunct}\relax
\EndOfBibitem
\bibitem[Nishiyama \latin{et~al.}(2003)Nishiyama, Katsura, Funakoshi, Kubo,
  Kubo, Tange, Sueda, and Yokoshi]{Nishiyama2003}
Nishiyama,~N.; Katsura,~T.; Funakoshi,~K.; Kubo,~A.; Kubo,~T.; Tange,~Y.;
  Sueda,~Y.; Yokoshi,~S. {Determination of the Phase Boundary between the
  \textit{B}1 and \textit{B}2 Phases in NaCl by \textit{in situ} X-ray
  Diffraction}. \emph{Phys. Rev. B} \textbf{2003}, \emph{68},
  134109--1--8\relax
\mciteBstWouldAddEndPuncttrue
\mciteSetBstMidEndSepPunct{\mcitedefaultmidpunct}
{\mcitedefaultendpunct}{\mcitedefaultseppunct}\relax
\EndOfBibitem
\bibitem[Matsuoka \latin{et~al.}(2019)Matsuoka, Hishida, Kuno, Hirao, Ohishi,
  Sasaki, Takahama, and Shimizu]{Matsuoka2018a}
Matsuoka,~T.; Hishida,~M.; Kuno,~K.; Hirao,~N.; Ohishi,~Y.; Sasaki,~S.;
  Takahama,~K.; Shimizu,~K. {Superconductivity of Platinum
  Hydride-Supplementary Material-}. \emph{Phys. Rev. B} \textbf{2019},
  \emph{99}, 144511--1--6\relax
\mciteBstWouldAddEndPuncttrue
\mciteSetBstMidEndSepPunct{\mcitedefaultmidpunct}
{\mcitedefaultendpunct}{\mcitedefaultseppunct}\relax
\EndOfBibitem
\bibitem[Zhang \latin{et~al.}(2013)Zhang, Oganov, Goncharov, Zhu, Boulfelfel,
  Lyakhov, Stavrou, Somayazulu, Prakapenka, and Konopkova]{Zhang}
Zhang,~W.; Oganov,~A.~R.; Goncharov,~A.~F.; Zhu,~Q.; Boulfelfel,~S.~E.;
  Lyakhov,~A.~O.; Stavrou,~E.; Somayazulu,~M.; Prakapenka,~V.~B.; Konopkova,~Z.
  {Unexpected Stable Stoichiometries of Sodium Chlorides}. \emph{Science}
  \textbf{2013}, \emph{342}, 1502--1505\relax
\mciteBstWouldAddEndPuncttrue
\mciteSetBstMidEndSepPunct{\mcitedefaultmidpunct}
{\mcitedefaultendpunct}{\mcitedefaultseppunct}\relax
\EndOfBibitem
\bibitem[Momma and Izumi(2011)Momma, and Izumi]{Momma2011}
Momma,~K.; Izumi,~F. {VESTA 3 for Three-Dimensional Visualization of Crystal,
  Volumetric and Morphology Data}. \emph{J. Appl. Crystallogr.}
  \textbf{2011}, \emph{44}, 1272--1276\relax
\mciteBstWouldAddEndPuncttrue
\mciteSetBstMidEndSepPunct{\mcitedefaultmidpunct}
{\mcitedefaultendpunct}{\mcitedefaultseppunct}\relax
\EndOfBibitem
\bibitem[Mao and Hemley(1994)Mao, and Hemley]{Mao1994a}
Mao,~H.-k.; Hemley,~R. {Ultrahigh-Pressure Transitions in Solid Hydrogen}.
  \emph{Rev. Mod. Phys.} \textbf{1994}, \emph{66}, 671--692\relax
\mciteBstWouldAddEndPuncttrue
\mciteSetBstMidEndSepPunct{\mcitedefaultmidpunct}
{\mcitedefaultendpunct}{\mcitedefaultseppunct}\relax
\EndOfBibitem
\bibitem[Boehler \latin{et~al.}(1997)Boehler, Ross, and Boercker]{Boehler1997}
Boehler,~R.; Ross,~M.; Boercker,~D.~B. {Melting of LiF and NaCl to 1 Mbar:
  Systematics of Ionic Solids at Extreme Conditions}. \emph{Phys. Rev.
  Lett.} \textbf{1997}, \emph{78}, 4589--4592\relax
\mciteBstWouldAddEndPuncttrue
\mciteSetBstMidEndSepPunct{\mcitedefaultmidpunct}
{\mcitedefaultendpunct}{\mcitedefaultseppunct}\relax
\EndOfBibitem
\bibitem[Dzyabura \latin{et~al.}(2013)Dzyabura, Zaghoo, and
  Silvera]{Dzyabura2013}
Dzyabura,~V.; Zaghoo,~M.; Silvera,~I.~F. {Evidence of a Liquid-Liquid Phase
  Transition in Hot Dense Hydrogen.} \emph{Proc. Natl. Acad. Sci. U. S. A.} \textbf{2013}, \emph{110},
  8040--1--4\relax
\mciteBstWouldAddEndPuncttrue
\mciteSetBstMidEndSepPunct{\mcitedefaultmidpunct}
{\mcitedefaultendpunct}{\mcitedefaultseppunct}\relax
\EndOfBibitem
\end{mcitethebibliography}

\begin{mcitethebibliography}{3}
\providecommand*\natexlab[1]{#1}
\providecommand*\mciteSetBstSublistMode[1]{}
\providecommand*\mciteSetBstMaxWidthForm[2]{}
\providecommand*\mciteBstWouldAddEndPuncttrue
  {\def\EndOfBibitem{\unskip.}}
\providecommand*\mciteBstWouldAddEndPunctfalse
  {\let\EndOfBibitem\relax}
\providecommand*\mciteSetBstMidEndSepPunct[3]{}
\providecommand*\mciteSetBstSublistLabelBeginEnd[3]{}
\providecommand*\EndOfBibitem{}
\mciteSetBstSublistMode{f}
\mciteSetBstMaxWidthForm{subitem}{(\alph{mcitesubitemcount})}
\mciteSetBstSublistLabelBeginEnd
  {\mcitemaxwidthsubitemform\space}
  {\relax}
  {\relax}

\bibitem[Zhang \latin{et~al.}(2013)Zhang, Oganov, Goncharov, Zhu, Boulfelfel,
  Lyakhov, Stavrou, Somayazulu, Prakapenka, and Konopkova]{Zhang}
Zhang,~W.; Oganov,~A.~R.; Goncharov,~A.~F.; Zhu,~Q.; Boulfelfel,~S.~E.;
  Lyakhov,~A.~O.; Stavrou,~E.; Somayazulu,~M.; Prakapenka,~V.~B.; Konopkova,~Z.
  {Unexpected Stable Stoichiometries of Sodium Chlorides}. \emph{Science}
  \textbf{2013}, \emph{342}, 1502--1505\relax
\mciteBstWouldAddEndPuncttrue
\mciteSetBstMidEndSepPunct{\mcitedefaultmidpunct}
{\mcitedefaultendpunct}{\mcitedefaultseppunct}\relax
\EndOfBibitem
\bibitem[Struzhkin \latin{et~al.}(2016)Struzhkin, Kim, Stavrou, Muramatsu, Mao,
  Pickard, Needs, Prakapenka, and Goncharov]{Struzhkin2016}
Struzhkin,~V.~V.; Kim,~D.~Y.; Stavrou,~E.; Muramatsu,~T.; Mao,~H.-k.;
  Pickard,~C.~J.; Needs,~R.~J.; Prakapenka,~V.~B.; Goncharov,~A.~F. {Synthesis
  of Sodium Polyhydrides at High Pressures}. \emph{Nat. Commun.}
  \textbf{2016}, \emph{7}, 12267--1--7\relax
\mciteBstWouldAddEndPuncttrue
\mciteSetBstMidEndSepPunct{\mcitedefaultmidpunct}
{\mcitedefaultendpunct}{\mcitedefaultseppunct}\relax
\EndOfBibitem
\end{mcitethebibliography}
\providecommand{\latin}[1]{#1}
\makeatletter
\providecommand{\doi}
  {\begingroup\let\do\@makeother\dospecials
  \catcode`\{=1 \catcode`\}=2 \doi@aux}
\providecommand{\doi@aux}[1]{\endgroup\texttt{#1}}
\makeatother
\providecommand*\mcitethebibliography{\thebibliography}
\csname @ifundefined\endcsname{endmcitethebibliography}
  {\let\endmcitethebibliography\endthebibliography}{}

\end{document}